\documentclass[preprint,showpacs,preprintnumbers,amsmath,amssymb,showkeys]{revtex4}

\usepackage{epsfig}
\usepackage{dcolumn}
\usepackage{bm}
\usepackage [T1]{fontenc}
\usepackage [latin1,utf8]{inputenc}
\usepackage{graphicx}
\usepackage{amsmath}
\usepackage{amsxtra}
\usepackage{color}
\usepackage[spanish,english]{babel}
\usepackage{wrapfig}
\usepackage{amssymb,amsmath}
\usepackage{makeidx}
\usepackage{graphics}
\usepackage{psfrag}
\usepackage{epsfig}
\usepackage{bm}
\usepackage{float}
\newcommand{\dbar} {\ensuremath{\, \mathchar'26\mkern-12mu d}}

\begin{document}

\title{Thermodynamic response functions and Maxwell relations for a Kerr black hole}

\author{L. Escamilla}
\altaffiliation[]{}
 \email{lescamilla@fisica.ugto.mx} 
\altaffiliation[]{}
\author{J. Torres-Arenas}
\email{jtorres@fisica.ugto.mx}
\affiliation{%
Divisi\'on de Ciencias e Ingenier\'ias Campus Le\'on, Universidad de Guanajuato, Loma del Bosque 103, 37150  Le\'on Guanajuato, M\'exico}%
\date{\today}

\begin{abstract}
Assuming the  existence of a fundamental thermodynamic relation, the classical thermodynamics of a black hole with mass and angular momentum is given.
 New definitions of response functions and $TdS$ equations are introduced and mathematical analogous of the Euler equation and Gibbs-Duhem relation are founded.
Thermodynamic stability is studied from concavity conditions, resulting in an unstable equilibrium at all the domain except for a small region of local stable equilibrium. Maxwell relations are written, allowing  to build the  thermodynamic squares. Our results shown an interesting analogy between thermodynamics of gravitational and magnetic systems. 
\end{abstract}


\keywords{Thermodynamics, Black Holes, Long Range Interactions}
\pacs{05.70.-a, 04.70.Bw, 95.30.Tg}

\maketitle

\section{Introduction}

Classical macroscopic thermodynamics is a branch of physics of great generality, applicable to systems of elaborate structure, with different physical properties, such as mechanical, thermal, electric and magnetic. Therefore, thermodynamics possess great importance in many fields of physics and technical sciences~\cite{Callen,Greiner}. The task of thermodynamics is to define suitable macroscopic quantities to describe average properties of systems and explain how are these quantities related by means of universally valid equations.

Even though thermodynamics is entirely of phenomenological nature, it is one of the best logically structured branches of physics. However, it is known, that systems which interact via the so-called long-range interactions, present non-conventional thermodynamic properties~\cite{Dauxois,Johal,Campa}. Gravitational, coulombic and dipolar systems, are some examples of long-range interactions, in which it is possible to observe non-common thermodynamic features, such as non-additivity, non-extensivity and the possibility of negative heat capacity. This work is inspired within this phenomenological thermodynamic spirit.

Bekenstein~\cite{Bekenstein1,Bekenstein2} proposed that an appropriate multiple of the event horizon area  must be interpreted as the entropy,
\begin{equation}\label{eq1}
S_{bh}=\frac{k_Bc^3}{4\hbar G}A;
\end{equation}
where $S_{bh}$ is the entropy of the black hole, $k_B$ is the Boltzmann constant, $c$ is the speed of light in vacuum, $\hbar$ is the Planck's constant divided by $2\pi$, $G$ is the universal constant of gravity  and $A$ is the area of the event horizon. 
Moreover, Hawking \emph{et al}~\cite{Hawking2} proved that certain laws of black hole mechanics are mathematical analogous of zeroth and first laws of thermodynamics.
Shortly after, it was proved that in addition to the laws of equilibrium thermodynamics, black holes obey as well, the standard theory of non-equilibrium thermodynamics in the form of a fluctuation-dissipation theorem~\cite{Candelas}.
Development of the so-called black hole thermodynamics still continues~\cite{Davies1,Wald, Martinez1,Martinez2,Custodio}.

Existence of non-zero temperature for a black hole supports the hypothesis that the relation between entropy and the area of the event horizon of a black hole has some physical significance. Knowledge of this quantity makes possible to obtain the entropy of a black hole as a function of some fundamental parameters which describes such system; namely, a thermodynamic fundamental equation for the black hole. 
Thermodynamics is used to find information about a gravitational system composed by a black hole, approaching to the problem from the classical thermodynamics point of view~\cite{Callen}.

Although a formal study of black hole thermodynamics can be explored from a more fundamental perspective~\cite{York,Braden,Brown,Lemos,Akbar}, it is not the goal of this work to pursue this path. Instead, in this work exploration of the thermodynamic information that can be extracted from the \textit{assumed} existence of a  thermodynamic fundamental equation is considered, as would be done for an ordinary thermodynamic system. This approach proves to be successful to recover the mathematical structure, completely analogous, to conventional systems, in particular magnetic ones. 

This paper is organized as follows. In section two, the fundamental thermodynamic equation of a black hole, its natural parameters and the derived equations of state are briefly discussed. The non-common features observed in these equations are examined, including how can they be interpreted. Mathematical analogous of the Euler equation and the Gibbs-Duhem relation for these systems are calculated. In section three new response functions are defined and analyzed, in particular, specific heat and its odd behavior is studied. With the aid of $TdS$ equations, independence of the new response functions are found. In section four, the three possible Legendre transformations are performed and the Maxwell relations are given. The mnemonic diagrams for those relations are constructed (thermodynamic squares) exhibiting an interesting relation with magnetic systems. A particular application of the formalism to the ``rotocaloric'' effect is discussed. In section five, stability conditions are reviewed and some consequences are analyzed. Some conclusions are given in section six.

\section{The fundamental equation: Gibbs-Duhem and Euler relations}

In this work, the simplest black holes, i.e., those with spherical symmetry~\cite{D'Inverno} are considered. The most general one is the Kerr-Newman black hole, which is characterized by three parameters: mass $M$, angular momentum $J$, and electric charge $Q$.
Fundamental equation in energy representation can be found by substituting the area $A(M,J,Q)$ in (\ref{eq1}), using $U=Mc^2$, and inverting the expression to get $U=U(S,J,Q)$:
\begin{equation}\label{eq3}
U=\Bigg\{\Bigg(\frac{\pi k_Bc^3}{\hbar GS}\Bigg)\Bigg[\Bigg(\frac{\hbar cS}{2\pi k_B}+\frac{Q^2}{2}\Bigg)^2+J^2c^2\Bigg]\Bigg\}^{1/2}.
\end{equation}
The first-order differential for the internal energy is:
\begin{equation}\label{eq6}
dU=\Bigg(\frac{\partial U}{\partial S}\Bigg)_{J,Q}dS+\Bigg(\frac{\partial U}{\partial J}\Bigg)_{S,Q}dJ+\Bigg(\frac{\partial U}{\partial Q}\Bigg)_{S,J}dQ.
\end{equation}

In analogy with internal energy of a thermodynamic simple system, $U=U(S,V,N_i)$, where $V$ stands for volume and $N_i$ for mole number of the $i$-th component, partial derivatives appearing in (\ref{eq6}) play the role of intensive parameters~\cite{Bekenstein1},

\begin{equation}\label{eq7}
T\equiv\Bigg(\frac{\partial U}{\partial S}\Bigg)_{J,Q}; \qquad \Omega\equiv\Bigg(\frac{\partial U}{\partial J}\Bigg)_{S,Q};\qquad  \Phi\equiv\Bigg(\frac{\partial U}{\partial Q}\Bigg)_{S,J}.
\end{equation}
Where $T$ is the temperature of the black hole, $\Omega$ its angular velocity and $\Phi$ describes the electric potential for a charge $Q$ evaluated at a distance equal to the event horizon radius~\cite{Hawking4}. 
These relations are equivalent to the EOS in a simple thermodynamic system.
Then, we have the expression:
\begin{equation}\label{eq8}
dU=TdS+\Omega dJ+\Phi dQ.
\end{equation}
Such relation is analogous to the thermodynamic fundamental equation of a simple system. In a conventional thermodynamic system, derivatives of internal energy with respect to extensive parameters give intensive parameters. 
This property is reflected in the fact that for a conventional thermodynamic system, the fundamental equation is a homogeneous first-order function of some extensive parameters. Similarly, the EOS of a conventional thermodynamic system are homogeneous zero order functions.
However, for the Kerr-Newman black hole, fundamental equation (\ref{eq3}), is not an homogeneous first-order function.
Two major implications for thermodynamics systems: first, it is an indication of non-additivity; second, non-extensivity is present, that implies that the EOS \textit{are not} homogeneous zero-order functions.

The fundamental cause of those particular thermodynamic properties lies in the gravitational nature of the system. Gravitation falls in the so-called long-range interactions. It is largely known that thermodynamical description of such systems presents several complications, among them, non-additivity and broken extensivity~\cite{Dauxois,Johal,Campa}.

For the Kerr-Newman black hole, EOS obtained by derivation of eq. (\ref{eq3}) are:
\begin{equation}\label{eq9}
\Omega=\Bigg(\frac{2\pi^2k_{B}^2c^5}{\hbar G}\frac{J}{S}\Bigg)\cdot\Sigma(S,J,Q),
\end{equation}

\begin{equation}\label{eq10}
\Phi=\Bigg(\frac{\pi k_Bc^4Q}{G}+\frac{\pi^2k_B^2c^3Q^3}{\hbar GS}\Bigg)\cdot\Sigma(S,J,Q),
\end{equation}

\begin{displaymath}
T=\Bigg(\frac{c^2}{2GS}\Bigg)\cdot\Sigma(S,J,Q)\Bigg\{c^2(\hbar cS+\pi k_BQ^2)
\end{displaymath}
\begin{equation}\label{eq11}
\qquad -\frac{c}{2\hbar S}[(\hbar cS+\pi k_BQ^2)^2+(2\pi k_BcJ)^2]\Bigg\};
\end{equation}
with $\Sigma(S,J,Q)$ given by:
\begin{equation}\label{eqn}
\Sigma=\Bigg\{\Bigg(\frac{\pi k_{B}c^3}{\hbar GS}\Bigg)\Big[(\hbar cS+\pi k_{B}Q^2)^2+(2\pi k_BcJ)^2\Big]\Bigg\}^{-1/2}.
\end{equation}
Given the astrophysical relevance of the angular momentum compared to the electric charge, henceforth in this work a Kerr black hole is considered, characterized by its mass $M$, and angular momentum $J$. As a consequence, a fundamental equation $U=U(S,J)$ by taking $Q=0$ in eq. (\ref{eq3}) is used, for the sake of simplicity.

We analyze (in natural units) the EOS, $T=T(S,J)$ and $\Omega=\Omega(S,J)$ by plotting each one with a fixed variable for each curve and studying its behavior. Starting with the angular velocity $\Omega(S,J)$, in FIG. 1, it is shown that for fixed entropy and increasing angular momentum, the angular velocity increases asymptotically until it reaches a maximum value. The rise of this asymptote can be understood by the physical limit imposed to a rotating black hole, which will be discussed later.\\ Considering FIG. 2, if the angular momentum is kept constant for an increasing entropy, the angular velocity decreases to zero asymptotically for maximum entropy. The system can produce large amounts of energy at the expense of its rotational energy, maximizing entropy~\cite{Bekenstein3}.\\
Similarly, for the temperature $T(S,J)$, there are two cases:  In FIG. 3, if $S$ is kept constant it is possible to notice that, for an increasing angular momentum, temperature falls rapidly to zero, therefore, the angular momentum plays the role of a temperature attenuator. During the process of energy production at expense of rotational energy, this energy is radiated as thermic radiation, reducing $J$ and increasing $T$.\\
Considering FIG. 4, for a fixed $J$, temperature reaches a maximum as entropy increases, in this plot there are two regions due to the competition between $J$ and $S$. There is a region where temperature is a monotonically increasing function of entropy, and a second region where $T$ drops asymptotically until it reaches a certain value for maximum entropy. Presence of those regions for temperature is related to the stability of the system. 

In the entropic representation, the system has a relation between internal energy and angular momentum, establishing a boundary beyond which the proposed model has no physical meaning. That happens since reality condition for entropy is violated in this limit,
\begin{equation}\label{eq12}
Gc^{-5}U^2\ge J.
\end{equation} 
This expression can be interpreted saying that for a fixed value of the internal energy, angular momentum has a limiting value, given by eq. (\ref{eq12}). Behavior of the angular momentum and the allowed region for internal energy are shown in FIG. 5. It is important to mention that this maximum value of $J$ also appears in general relativity, referring to extreme Kerr black holes and naked singularities~\cite{Misner}. Existence of the same prediction in two different approaches is a good indication that postulation of a fundamental thermodynamic equation reasonably captures the essence of the system.

Following the canonical thermodynamic formalism, analogous of Euler equation, since extensivity is broken, is calculated. The internal energy is an homogeneous $1/2$-order function, which implies that:
\begin{equation}\label{eq13}
2TS+2\Omega J=U.
\end{equation}
From this expression, a Gibbs-Duhem-like relation can be calculated,
\begin{equation}\label{eq14}
dU=-2(SdT+Jd\Omega);
\end{equation} 
with this procedure, a relation for energy as a function of temperature and angular velocity $U=U(T,\Omega)$ is obtained. However, eq. (\ref{eq13}) it is not a fundamental thermodynamic equation, for this function it follows:

\begin{equation}\label{eq14+1}
\frac{\partial U}{\partial T}=-2S; \qquad \frac{\partial U}{\partial\Omega}=-2J.
\end{equation}

\section{Response functions and $TdS$ equations}

Second derivatives of the fundamental equation, also called \textit{response functions}, are usually descriptive of material properties and often these quantities are of the most direct physical interest. For the Kerr black hole we proposed the following basic set of response functions, composed by $\alpha_{\Omega},\chi_T,C_\Omega$. With this particular set, it is possible to recover all the thermodynamic information. As we will show, definitions given below reproduces the conventional structure of thermodynamics, showing a close analogy with the thermodynamics of magnetic systems but different to the thermodynamics of fluids. To our knowledge, description of thermodynamics for the Kerr black hole, using this set of response functions, it is presented for the first time.

Coefficient $\alpha_{\Omega}$ is defined by:

\begin{equation}\label{eq15+1}
\alpha_{\Omega}\equiv\Bigg(\frac{\partial J}{\partial T}\Bigg)_{\Omega}.
\end{equation}
The quantity $\alpha_{\Omega}$ is the change in the angular momentum per unit change in the temperature of a Kerr black hole maintained at constant angular velocity.\\
The response function $\chi_T$ is given by:

\begin{equation}\label{eq15+2}
\chi_T\equiv\Bigg(\frac{\partial J}{\partial\Omega}\Bigg)_T.
\end{equation} 
$\chi_T$ is the increase in the angular momentum per unit change in the angular velocity at constant temperature, and can be interpreted as an isothermic rotational susceptibility.\\
The heat capacity at constant angular velocity is defined by:

\begin{equation}\label{eq15+3}
C_\Omega\equiv T\Bigg(\frac{\partial S}{\partial T}\Bigg)_\Omega=\Bigg(\frac{\dbar Q}{dT}\Bigg)_{\Omega}.
\end{equation}
The heat capacity at constant angular velocity is the quasi-static heat flux required to produce unit change in the temperature of a Kerr black hole maintained at constant angular velocity.

All the other response functions of the Kerr black hole can be expressed in terms of this basic set $C_\Omega,\chi_T,\alpha_\Omega$. This set of functions are in fact, the three possible second derivatives of the thermodynamic potential $G=G(T,\Omega)$ that will be introduced in section three,

\begin{equation}\label{eq15+4}
\chi_T=-\Bigg(\frac{\partial^2 G}{\partial\Omega^2}\Bigg)_T;  \quad \alpha_{\Omega}=-\Bigg(\frac{\partial^2 G}{\partial T\partial\Omega}\Bigg); \quad C_{\Omega}=-T\Bigg(\frac{\partial^2 G}{\partial T^2}\Bigg)_{\Omega}.
\end{equation}
Two additional response functions of interest are the heat capacity at constant angular momentum $C_J$, and the isentropic susceptibility, $\chi_S$. \\
The heat capacity at constant angular momentum, defined by:

\begin{equation}\label{eq15+5}
C_J\equiv T\Bigg(\frac{\partial S}{\partial T}\Bigg)_J=\Bigg(\frac{\dbar Q}{dT}\Bigg)_J,
\end{equation}
it is a measure of the quasi-static heat flux needed to change in one unit the temperature of a Kerr black hole maintained at constant angular momentum. \\
The response function $\chi_S$ is given by:

\begin{equation}\label{eq15+6}
\chi_S\equiv\Bigg(\frac{\partial J}{\partial \Omega}\Bigg)_S.
\end{equation}
This quantity characterizes the change in the angular momentum associated with an isentropic change in the angular velocity.

The $TdS$ equations are necessary to obtain relations between response functions. These equations can be found by considering entropy as a function of certain natural variables, differentiating and rearranging the expression to find a relation between $TdS$ and certain second derivatives.  
For a Kerr black hole the first $TdS$ equation can be found by considering entropy as a function of the temperature and the angular velocity, $S=S(T,\Omega)$, 

\begin{equation}\label{eq15+6+1}
TdS=C_\Omega dT+T\Bigg(\frac{\partial J}{\partial T}\Bigg)_\Omega d\Omega;
\end{equation}
the second $TdS$ equation can be obtained considering entropy as a function of the temperature and the angular momentum, $S=S(T,J)$,

\begin{equation}\label{eq15+6+2}
TdS=C_JdT-T\Bigg(\frac{\partial \Omega}{\partial T}\Bigg)_J dJ.
\end{equation}
Relations between response functions can be found through manipulation of $TdS$ equations and the use of Maxwell relations, discussed in section four. Two relations that describe the dependence of $C_J$ and $\chi_S$ in terms of the basic set of response functions are:

\begin{equation}\label{eq15+7}
\chi_T(C_{\Omega}-C_{J})=T\alpha^2_{\Omega},
\end{equation}
\begin{equation}\label{eq15+8}
C_{\Omega}(\chi_T-\chi_S)=T\alpha^2_{\Omega}.
\end{equation}

The specific heat at constant angular momentum is calculated,

\begin{equation}\label{eq15}
C_J=\Bigg(\frac{\partial U}{\partial T}\Bigg)_J=-\frac{1}{2T^2}\Bigg(\frac{\hbar c^5}{4\pi k_BG}-\frac{\pi k_Bc^5J^2}{\hbar GS^2}\Bigg).
\end{equation}
It is possible to notice the existence of a negative region. Since the second term is reduced for an increasing entropy and subtraction inside the brackets will eventually result in a positive quantity, causing that the specific heat becomes negative, for sufficiently large values of entropy.
Negative region for the  specific heat is known long ago for gravitational systems~\cite{Padmanabhan} in general, and for black holes in particular~\cite{Dauxois,Campa,Davies1,Oppenheim}.
Two interpretations about negative specific heat arise. \\
One is related to the transfer of energy as heat; having a region of negative heat implies that the temperature of a black hole raises to a higher value, with energy transfer~\cite{Parker}. The second one is related to thermodynamic stability of the system. For a negative specific heat, thermodynamic systems are in unstable states passing perhaps, through a phase transition~\cite{Davies2}.

\section{Thermodynamic potentials and Maxwell relations}

Another point of interest are the thermodynamic potentials. Alternative representations for the thermodynamic fundamental equation, can be obtained via the application of a Legendre transformation to $U=U(S,J)$. For a Kerr black hole there are three possible Legendre transformation. 

Starting with the mathematical analogous of the \textit{Helmholtz free energy}, $F=F(T,J)=U[T]$, 

\begin{equation}\label{eq16}
F=U-TS, \qquad \textrm{and} \qquad -S=\frac{\partial F}{\partial T};
\end{equation}
which can be written as:

\begin{equation}\label{eq17}
F=\Sigma(S,J)\cdot\Bigg[\frac{1}{2}\Bigg(\frac{\hbar c^5}{4\pi k_BG}\Bigg)S+\frac{3J^2}{2S}\Bigg(\frac{\hbar c^3}{2\pi k_B}\Bigg)\Bigg];
\end{equation}
with $\Sigma(S,J)$ given by eq. (\ref{eqn}), for $Q=0$. It is worth to mention that eq. (\ref{eq17}) must be a function of temperature and angular momentum as independent variables; instead, eq. (\ref{eq17}) is a function of the entropy and the angular momentum, $F=F(S,J)$. We were not able to find an analytical expression for Helmholtz free energy in that terms.\\ 
The  Helmholtz free energy for a Kerr black hole in diathermal contact with a heat reservoir can be interpreted as the available work at a constant temperature, since $dW=\Omega dJ$. 

A second Legendre transformation, can be obtained replacing the angular momentum $J$, by the angular velocity $\Omega$, as the independent variable in fundamental equation, $H=H(S,\Omega)=U[\Omega]$, this leads to:

\begin{equation}\label{eq18}
H=U-\Omega J, \qquad \textrm{and} \qquad -J=\frac{\partial H}{\partial\Omega};
\end{equation}
the potential $H$ is given by:

\begin{equation}\label{eq19}
H=\sqrt{\frac{c^4S}{2G}}\cdot\Bigg[1+\Bigg(\frac{\Omega^2}{\frac{k_Bc^5}{\hbar GS}-\Omega^2}\Bigg)\Bigg]^{1/2};
\end{equation}
which is a function of entropy and angular velocity as independent variables. Therefore, eq. (\ref{eq19}) is an explicit expression for the potential $H=H(S,\Omega)$. It can be interpreted as the heat added to  a Kerr black hole
at constant  angular velocity, since $dQ=TdS$.

The last thermodynamic potential is obtained by simultaneously replacing both, the entropy and the angular momentum, by the temperature and the angular velocity as independent variables in the fundamental equation,  $G=G(T,\Omega)=U[T,\Omega]$,

\begin{equation}\label{eq20}
G=U-TS-\Omega J, \qquad \textrm{and} \qquad -J=\frac{\partial G}{\partial\Omega}, \qquad -S=\frac{\partial G}{\partial T};
\end{equation}
an expression for the $G$ potential is given by:

\begin{equation}\label{eq21}
G=\frac{1}{2}\Bigg[\Bigg(\frac{\pi k_Bc^3}{\hbar GS}\Bigg)\Bigg(\frac{\hbar^2c^2}{4\pi^2k_B^2}S^2+J^2c^2\Bigg)\Bigg]^{1/2}.
\end{equation}
As for Helmholtz free energy in eq. (\ref{eq17}), an expression  for $G(T,\Omega)$ as a function of $S$ and $J$ instead of its natural variables was obtained. Physical interpretation of this potential for a Kerr black hole it is not clear, since in a conventional thermodynamic system, the mathematical analogous of this potential used to be associated with a third extensive parameter, missing in this case. However, for a Kerr-Newman black hole, this third parameter is the electric charge $Q$, therefore, this potential can be interpreted as the available work at constant temperature and angular velocity, since $dW=\Phi dQ$, for a Kerr-Newman black hole.   

Different representations for the fundamental equation leads to the existence of various derivatives of independent variables related to each thermodynamic potential, apparently unrelated with other derivatives.  However, there are some relations among such derivatives; hence, just a few of them can be considered as independent. These relations between derivatives are known as Maxwell relations and arise from the equality of the mixed partial derivatives of the fundamental equation, expressed in any of its alternative representations. 
Since there are only two independent variables, no more than a pair of mixed derivatives exists in any representation of the fundamental equation. For the energetic representation $U=U(S,J)$, Maxwell relations are:
\begin{equation}\label{eq22}
(S,\Omega) \qquad \frac{\partial^2 H}{\partial S\partial \Omega}=\frac{\partial^2 H}{\partial\Omega\partial S} \Rightarrow -\Bigg(\frac{\partial J}{\partial S}\Bigg)_\Omega=\Bigg(\frac{\partial T}{\partial \Omega}\Bigg)_S,
\end{equation}
\begin{equation}\label{eq23}
(T,\Omega) \qquad \frac{\partial^2 G}{\partial T\partial\Omega}=\frac{\partial^2 G}{\partial \Omega\partial T} \Rightarrow \Bigg(\frac{\partial J}{\partial T}\Bigg)_\Omega=\Bigg(\frac{\partial S}{\partial \Omega}\Bigg)_T,
\end{equation}
\begin{equation}\label{eq24}
(T,J) \qquad \frac{\partial^2 F}{\partial J\partial T}=\frac{\partial^2 F}{\partial T\partial J} \Rightarrow -\Bigg(\frac{\partial S}{\partial J}\Bigg)_T=\Bigg(\frac{\partial \Omega}{\partial T}\Bigg)_J,
\end{equation}
\begin{equation}\label{eq25}
(S,J) \qquad \frac{\partial^2 U}{\partial J\partial S}=\frac{\partial^2 U}{\partial S\partial J} \Rightarrow \Bigg(\frac{\partial T}{\partial J}\Bigg)_S=\Bigg(\frac{\partial \Omega}{\partial S}\Bigg)_J.
\end{equation}

Maxwell relations can be remembered conveniently in terms of simple mnemonic diagrams~\cite{Koenig1}. In this case, the construction rules of those diagrams are inspired in the construction of the same mnemonic diagrams for magnetic systems~\cite{Stanley}:  A minus sign is added to the extensive variable of magnetic work, with this adaptation it is possible to build the mnemonic diagrams for magnetic systems. For Kerr black holes, the same result can be achieved by adding a minus sign to the angular momentum. Maxwell relations are represented in the next diagrams.
In FIG. 7, the Maxwell relation for the potential $H$, (\ref{eq22}) is depicted. The remaining Maxwell relations are represented in FIG. 8 - FIG. 10 as successive rotations of the thermodynamic square shown in FIG. 7. 

\subsection{The ``Rotocaloric Effect''}

An additional example of the similarities between thermodynamics of magnetic systems and thermodynamics of black holes, lies in the exploration of the magnetocaloric effect (MCE) and its counterpart for a Kerr black hole, here named as ``rotocaloric effect''.

The MCE is a magnetothermal phenomena exhibited by magnetic materials, in which a reversible change in temperature of the material occurs under adiabatic and usually isobaric conditions in the presence of a variable magnetic field~\cite{Tishin}. For a magnetic system with a total differential for entropy expressed as a function of temperature $T$, and the magnetic field $H$, the change in temperature under variations of $H$, can be obtained from:

\begin{equation}\label{eq_rot1}
\frac{dT}{dH}=-\frac{(\partial S/\partial H)_T}{(\partial S/\partial T)_H}.
\end{equation}

For a Kerr black hole a similar process could appear, a reversible change in temperature of the event horizon under adiabatic conditions due to variations in the angular momentum of the black hole. To calculate this changes in $T$, the $TdS$ equation (\ref{eq15+6+2}), is used. In an adiabatic process $TdS=0$:

\begin{equation}\label{eq_rot2}
dT=\frac{T}{C_J}\Big(\frac{\partial\Omega}{\partial T}\Big)_J dJ;
\end{equation}
using the definition of $C_J$ and the Maxwell relations, it is possible to express the variation in $T$ as:

\begin{equation}\label{eq_rot3}
\frac{dT}{dJ}=-\frac{(\partial S/\partial J)_T}{(\partial S/\partial T)_J}.
\end{equation}

To compare both phenomena, the MCE is applied to a paramagnetic substance composed by $N$ freely removable dipoles with a permanent magnetic dipole 
moment $\vec{\mu}$~\cite{Greiner}. Considering only the dominant term in entropy $S(T,H)$, the variation in the temperature of this paramagnetic substance under changes in the applied magnetic field, is given by:

\begin{equation}\label{eq_rot6}
\frac{dT}{dH}=\frac{T}{H}.
\end{equation}

Similarly, to calculate the value of the rotocaloric effect for a Kerr black hole, an explicit relation for entropy as a function of temperature and angular momentum is required. The equation of state for temperature eq. (\ref{eq11}), considering $Q=0$, was used to find this relation. It is necessary to invert $T=T(S,J)$ to obtain $S=S(T,J)$. An approximated relation $S(T,J)$, was estimated considering an iterative process.\\ For a Kerr black hole, the first-order approximation was calculated proposing $S_0$ as the entropy of a Schwarzschild black hole, the simplest one, by considering $J=Q=0$ in eq. (\ref{eq3}), and taking  the derivative with respect of temperature: 

\begin{equation}\label{eq_rot7}
S_0(T)=\frac{\hbar c^5}{16\pi Gk_B}\frac{1}{T^2}.
\end{equation}
With this relation, the first-order approximation $S_1(T,J)$, for the exact relation $S(T,J)$, is given by:

\begin{equation}\label{eq_rot8}
S_1=\frac{\gamma_1}{J^4T^{10}}+\frac{\gamma_2}{J^2T^6};
\end{equation}
where $\gamma_1$ and $\gamma_2$ are a combination of the fundamental constants used in this work. If only the dominant term $1/J^2T^6$ is considered in eq. (\ref{eq_rot3}), the variation in the temperature under changes in the angular momentum for a Kerr black hole is:

\begin{equation}\label{eq_rot9}
\frac{dT}{dJ}=-\frac{T}{3J}.
\end{equation}
As $J$ and $T$ are non-negative in all the entropy domain, the rotocaloric effect produces a drop in the temperature of the black hole for an increase in the angular momentum and vice versa, this result was previously obtained analyzing the equation of state for the angular velocity, eq. (\ref{eq9}) in section two. 
The results in eqs. (\ref{eq_rot6}) and (\ref{eq_rot9}) are very similar, but with different sign. Therefore the changes in temperature work in the opposite direction for rotocaloric effect, compared with magnetic systems. 

Due the analogies between MCE and rotocaloric effect, it is possible to construct a refrigeration cycle for the Kerr black hole. This process is very similar than the magnetic one, only with the opposite sign.\\ A deeper exploration of these effects will be given in a subsequent work.

\section{Stability conditions}

As is discussed in section three, exist a negative region for the specific heat $C_J$, which indicates non-stability in the system. It is convenient to use the entropic representation of the thermodynamic fundamental equation to analyze stability conditions,

\begin{equation}\label{eq26}
S=\frac{2\pi k_B}{\hbar c}\Bigg[Gc^{-4}U^2+\Big(Gc^{-8}U^4-c^2J^2\Big)^{1/2}\Bigg].
\end{equation} 
The system is thermodynamically stable one, if the entropy $S(U,J)$ satisfies:

\begin{equation}\label{eq27}
S(U,J+\Delta J)+S(U,J-\Delta J)\leq2S(U,J),
\end{equation}
\begin{equation}\label{eq28}
S(U+\Delta U,J)+S(U-\Delta U,J)\leq2S(U,J);
\end{equation}
or for infinitesimal displacements:

\begin{equation}\label{29}
\Bigg(\frac{\partial^2S}{\partial U^2}\Bigg)_J\leq0, \qquad \Bigg(\frac{\partial^2S}{\partial J^2}\Bigg)_U\leq0.
\end{equation}
For a stable system, the graphical representation of the previous conditions results in a concave curve for entropy. If the entropy has a non-concave curve, the related system is unstable. When the entropy possess both, concave and non-concave regions, the system is not stable; instead, it is a locally stable one, but globally unstable.

Stability under variations in the angular momentum, at a fixed internal energy, is analyzed in FIG. 11. Entropy is a concave function in all its range of definition for the exchange of angular momentum at constant internal energy. Therefore, the Kerr black hole is stable under changes in the angular momentum.\\
If the angular momentum is kept constant and variations of the internal energy are allowed,  it is possible to notice that the entropy is non-concave in almost all its range, except for a region of very low energy, where the entropy is a concave function. Hence, a Kerr black hole is locally stable around $U\approx0$, but it is globally unstable under variations in its internal energy, as can be seen in FIG. 12.

If variations in both, the angular momentum and the internal energy are allowed, the three-dimensional space $(S,U,J)$, must be considered. The global stability condition requires that the entropy surface $S(U,J)$ lies everywhere below its tangent planes. Accordingly, for arbitrary $\Delta U$ and $\Delta J$:

\begin{equation}\label{eq30}
S(U+\Delta U,J+\Delta J)+S(U-\Delta U,J-\Delta J)\le2S(U,J);
\end{equation}
and locally, for $\Delta U\to0$ $\Delta U\to0$:

\begin{equation}\label{eq31}
\frac{\partial^2 S}{\partial U^2}\frac{\partial^2 S}{\partial J^2}-\Bigg(\frac{\partial^2S}{\partial U\partial J}\Bigg)\ge0.
\end{equation}
Behavior of the entropy in the subspace $(S,U,J)$ for a Kerr black hole, is shown in FIG. 13. Entropy is a non-concave function of $U$ and $J$; in correspondence with two-dimensional case for the entropy as a function of the internal energy, a region of stability exists. This locally stable region is available for small values of energy and angular momentum. \\
Evidence that a Kerr black hole is a thermodynamic system in unstable equilibrium in general, but in a state of locally stable equilibrium for very small values of its mass and angular momentum, is founded.

\section{Conclusions}

The implementation of the conventional thermodynamic formalism for black holes is permitted by the postulation of a fundamental thermodynamic equation of the form $U=U(S,J,Q)$. The proposal of considering a purely thermodynamic vision appears to be fruitful.
The application of this formalism in such systems presents particular complications, as non-additivity or instability of equilibrium states, typical of a system with dominant gravitational interactions; however, this development shows the presence of considerations suggesting that, beyond the mathematical exercise, the proposed model really represents some aspects of a very interesting behavior. This can be seen in the predictions made, that can be found in other frameworks of physics and in resemble with the thermodynamics of magnetic systems. The presented formalism could be a start point for the research of interesting analogous thermodynamic processes for black holes, as the Joule-Thomson effect and others. Results about these topics will be presented in a future work.

\section*{Acknowledgments}

We want to acknoweledge CONACyT (proyect-152684) and the University of Guanajuato (DAIP-006/10) for the support in the realization of this work.

\newpage

\section*{Figure captions}

\textbf{Figure 1.} Angular velocity vs. angular momentum for a given entropy. Asymptote in $\Omega$ is linked to the so-called Extreme Kerr black hole.\\

\textbf{Figure 2.} Angular velocity at a given $J$. Rotation rate decreases as the thermal energy is radiated increasing entropy.\\

\textbf{Figure 3.} Temperature at fixed $S$. Inverse function $J=J(T)$ is plotted to show that $T$ increases when the rotating black hole transforms its rotational energy into thermal radiation.\\

\textbf{Figure 4.} Temperature at given $J$. The maximum  is related with stability and the sign of the specific heat.\\

\textbf{Figure 5.} Forbidden region (over the curve) for angular momentum.\\

\textbf{Figure 6.} Specific heat $C_J$ showing negative values for large enough entropy.\\

\textbf{Figure 7.} Thermodynamic square representing Maxwell relation (\ref{eq22}) for the thermodynamic potential $H=H(S,\Omega)$.\\

\textbf{Figure 8.} Thermodynamic square representing Maxwell relation (\ref{eq23}) for the thermodynamic potential $G=G(T,\Omega)$.\\

\textbf{Figure 9.} Thermodynamic square of Maxwell relation (\ref{eq24}) for Helmholtz free energy, $F=F(T,J)$.\\

\textbf{Figure 10.} Thermodynamic square representation of Maxwell relation (\ref{eq25})  for internal energy, $U=U(S,J)$.\\

\textbf{Figure 11.} Entropic representation of fundamental equation $S(U,J)$ at constant energy. Concavity of the curve implies  stability.\\

\textbf{Figure 12.} Entropy $S(U,J)$ at a fixed $J$. Entropy is non-concave at almost all their domain, except for a small region of low energy.\\

\textbf{Figure 13.} Three-dimensional space $(S,U,J)$, showing unstable equilibrium for the system.\\

\newpage

\section*{Figures}

\begin{figure}[H]
\centering
\includegraphics[width= 10.0cm]{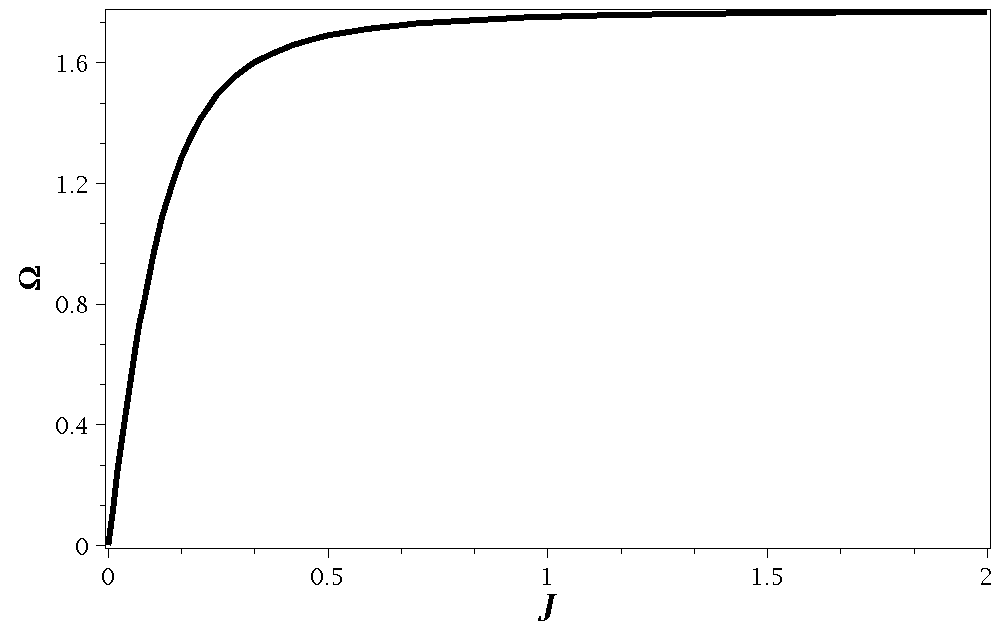} \\
\label{fig1}
\end{figure}
Figure 1.\\

\begin{figure}[H]
\centering
\includegraphics[width= 10.0cm]{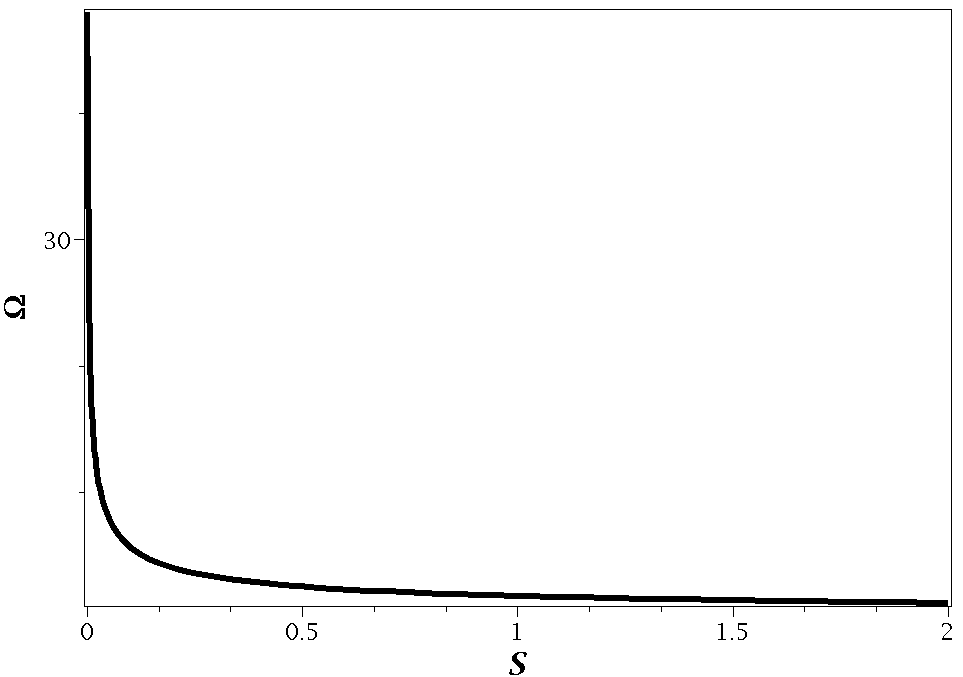} \\
\label{fig2}
\end{figure}
Figure 2.\\

\begin{figure}[H]
\centering
\includegraphics[width= 10.0cm]{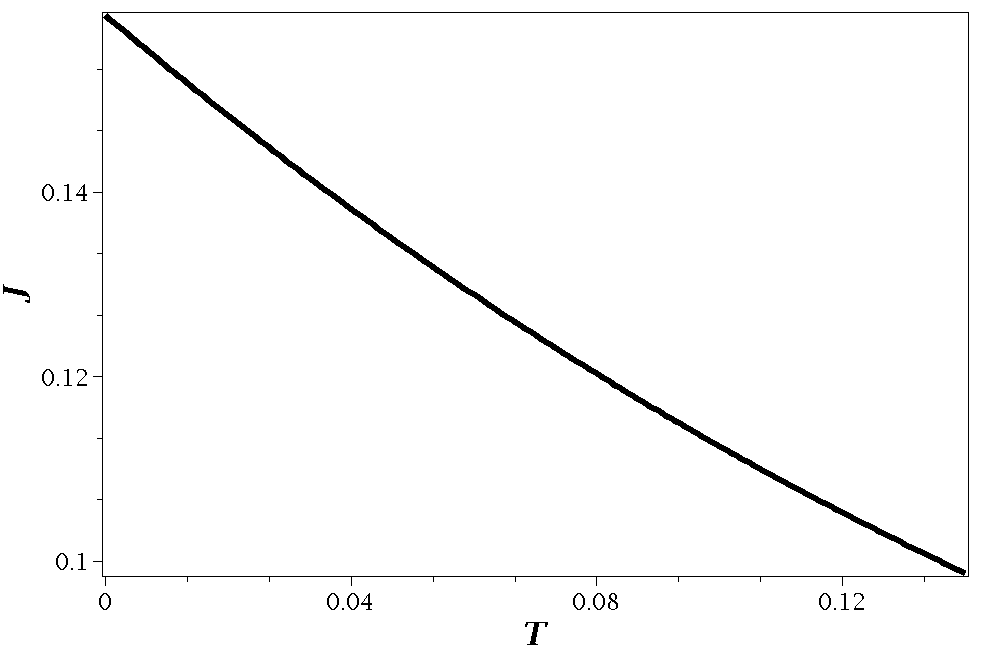} \\
\label{fig3}
\end{figure}
Figure 3.\\

\begin{figure}[H]
\centering
\includegraphics[width= 10.0cm]{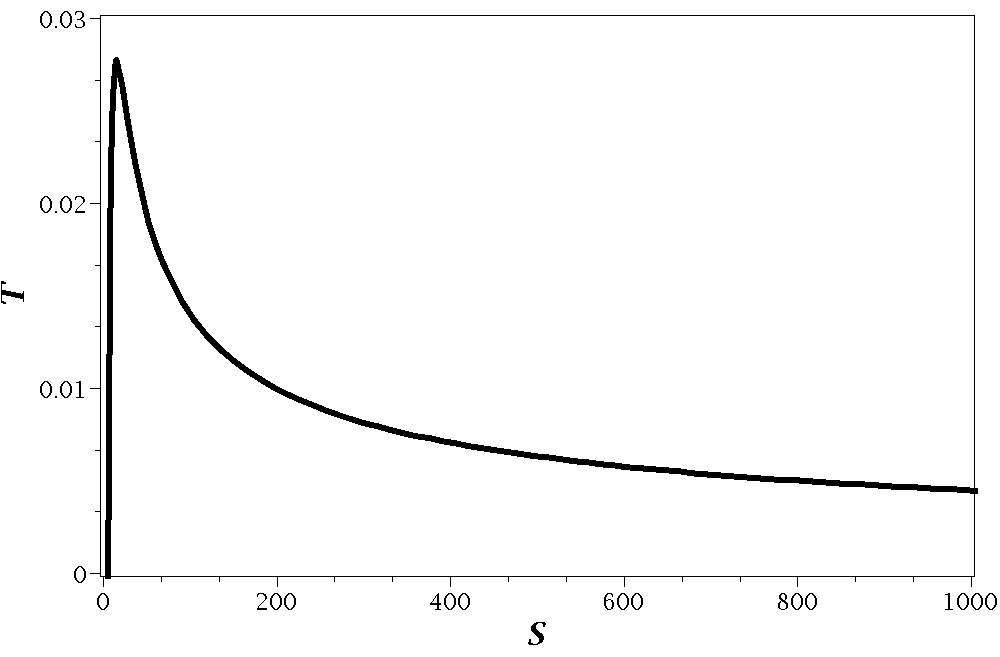} \\
\label{fig4}
\end{figure}
Figure 4.\\

\begin{figure}[H]
\centering
\includegraphics[width= 10.0cm]{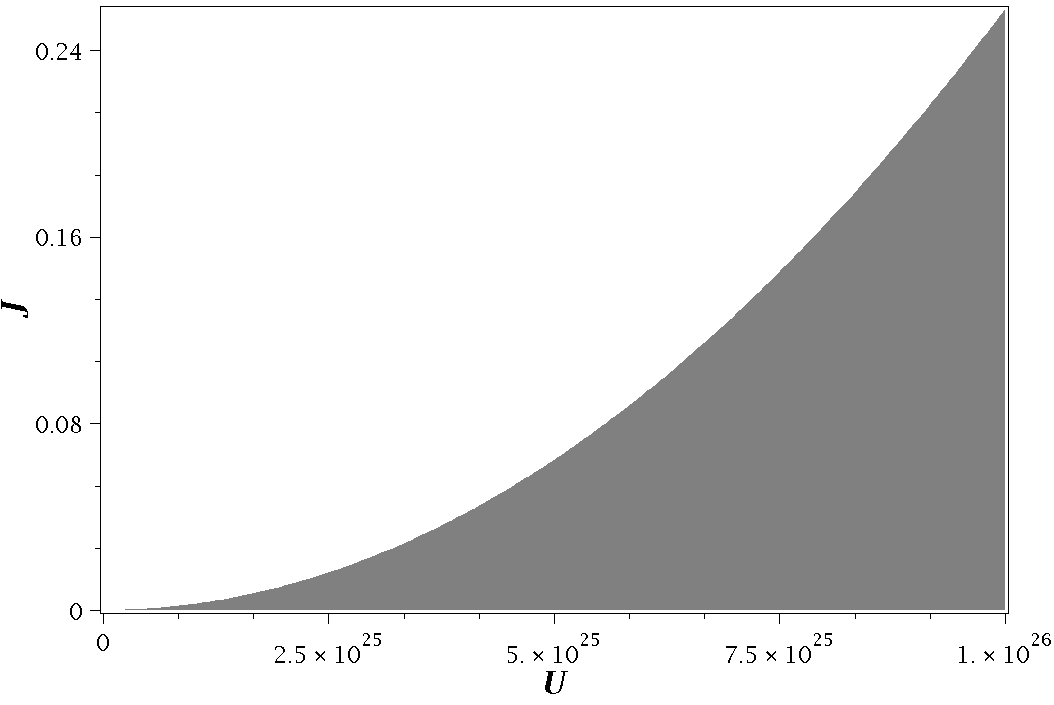} \\
\label{fig5}
\end{figure}
Figure 5.\\

\begin{figure}[H]
\centering
\includegraphics[width= 10.0cm]{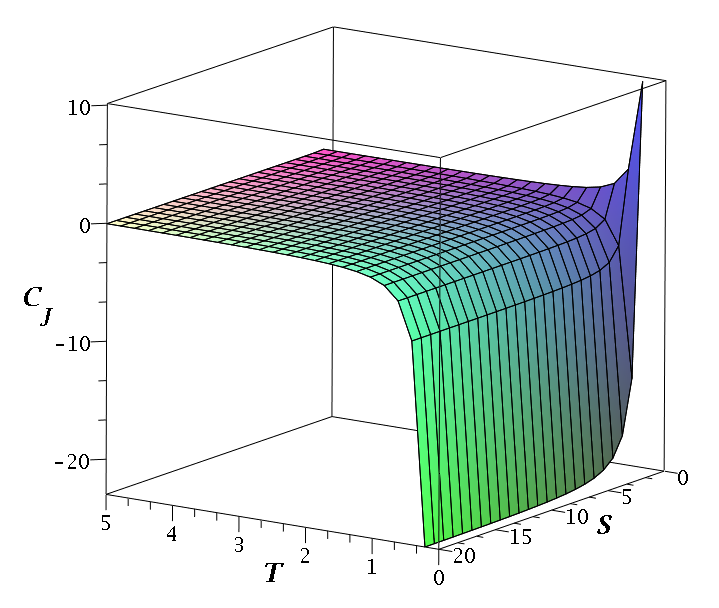} \\
\label{fig6}
\end{figure}
Figure 6.\\

\begin{figure}[H]
\centering
\includegraphics[width= 8.0cm]{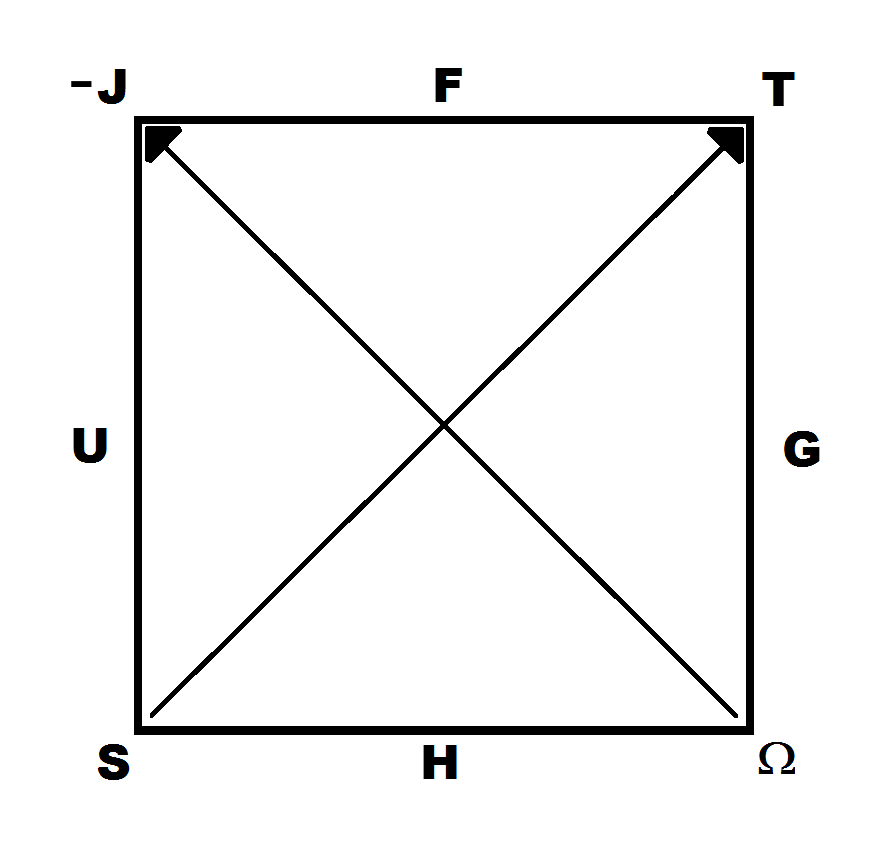} \\
\label{fig7}
\end{figure}
Figure 7.\\

\begin{figure}[H]
\centering
\includegraphics[width= 8.0cm]{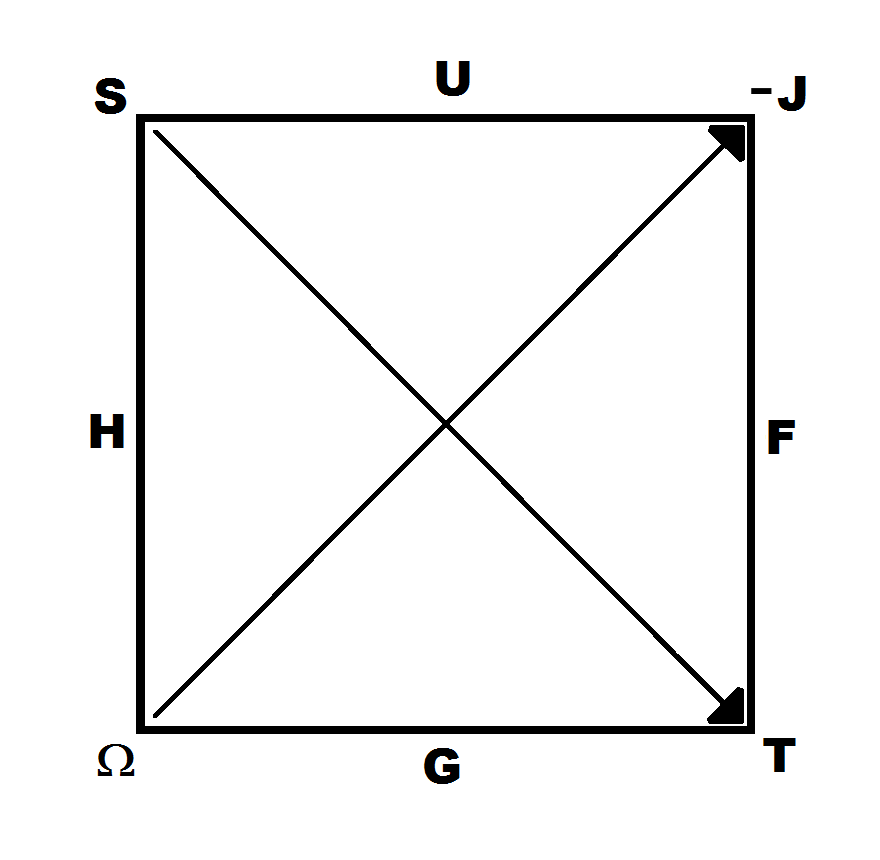} \\
\label{fig8}
\end{figure}
Figure 8.\\

\begin{figure}[H]
\centering
\includegraphics[width= 8.0cm]{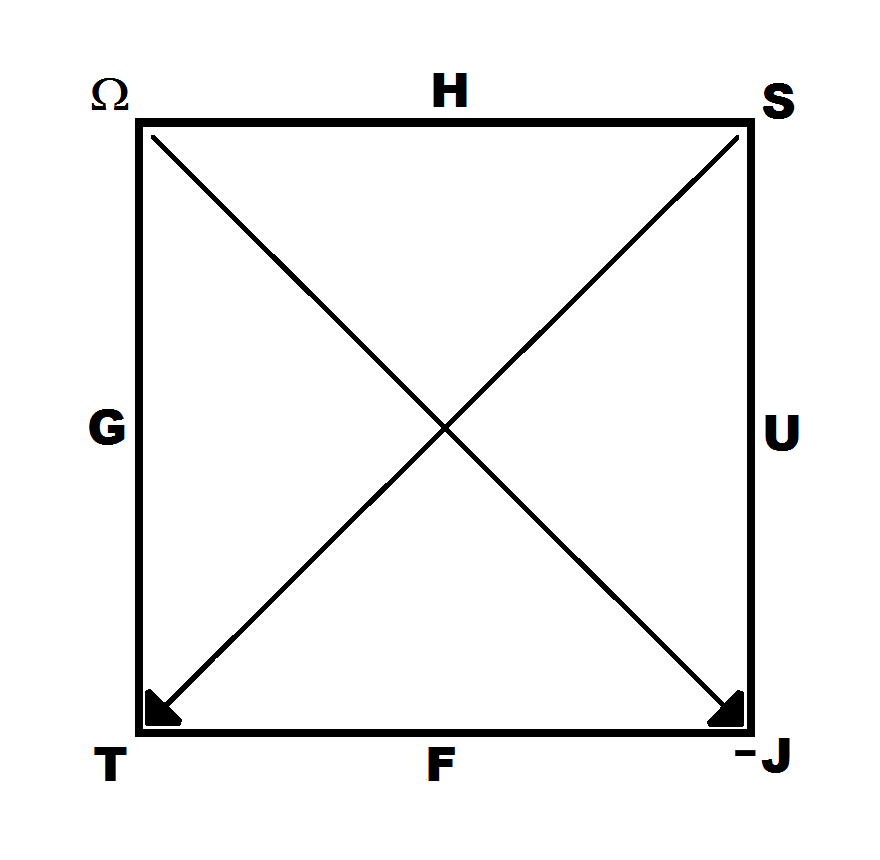} \\
\label{fig9}
\end{figure}
Figure 9.\\

\begin{figure}[H]
\centering
\includegraphics[width= 8.0cm]{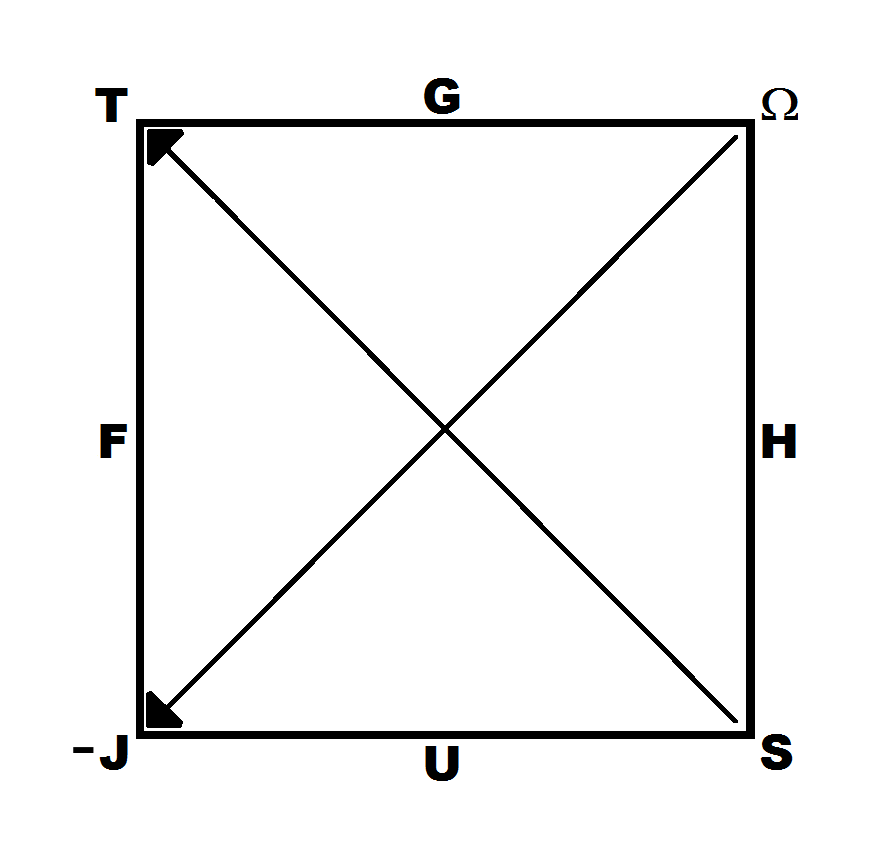} \\
\label{fig10}
\end{figure}
Figure 10.\\

\begin{figure}[H]
\centering
\includegraphics[width= 10.0cm]{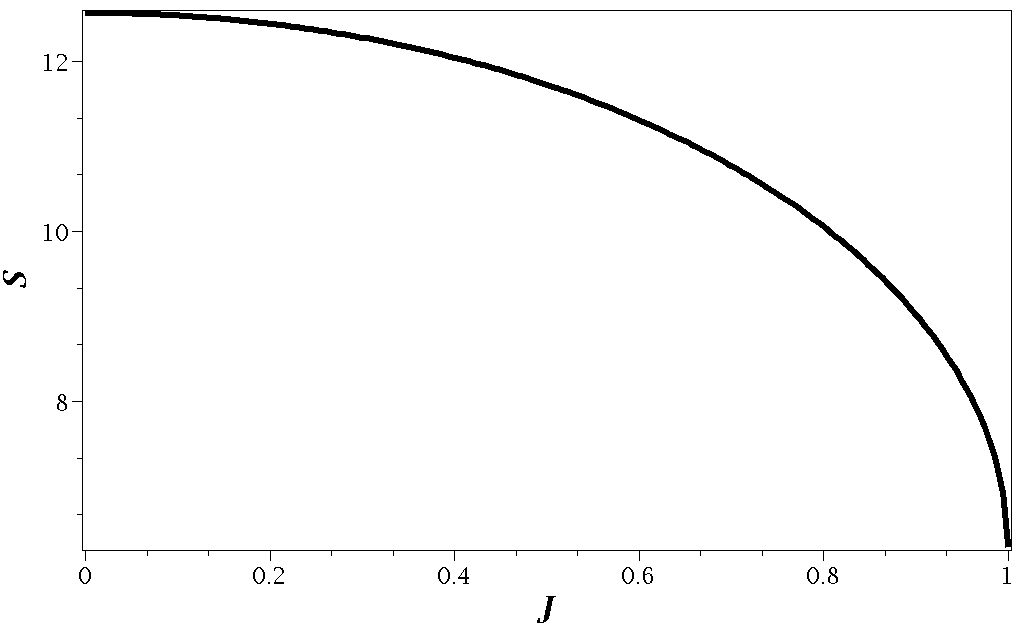} \\
\label{fig11}
\end{figure}
Figure 11.\\

\begin{figure}[H]
\centering
\includegraphics[width= 10.0cm]{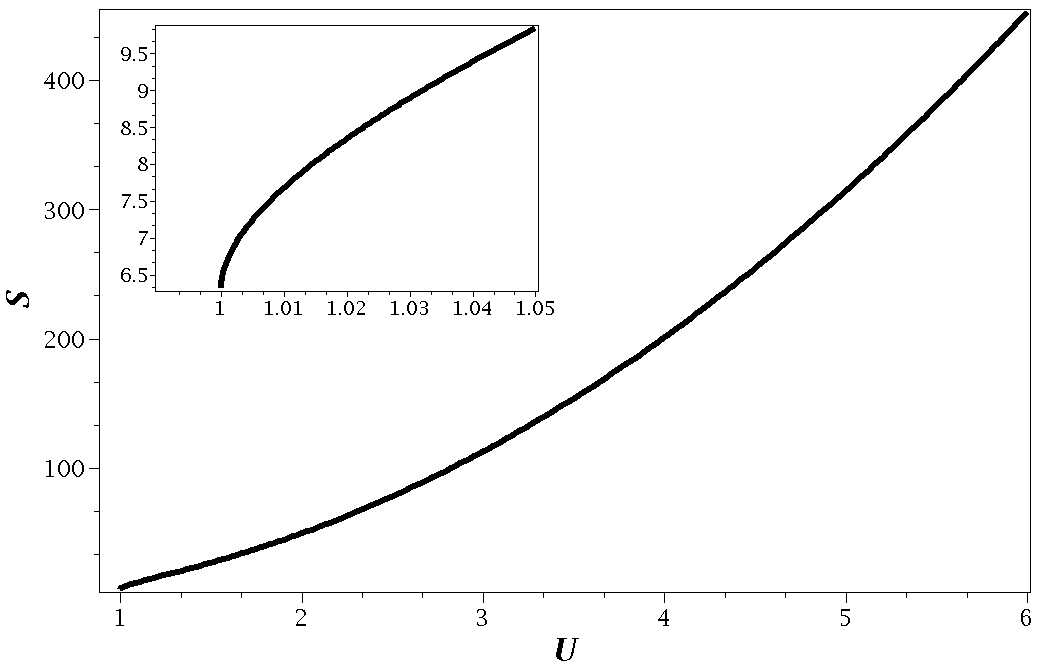} \\
\label{fig12}
\end{figure}
Figure 12.\\

\begin{figure}[H]
\centering
\includegraphics[width= 10.0cm]{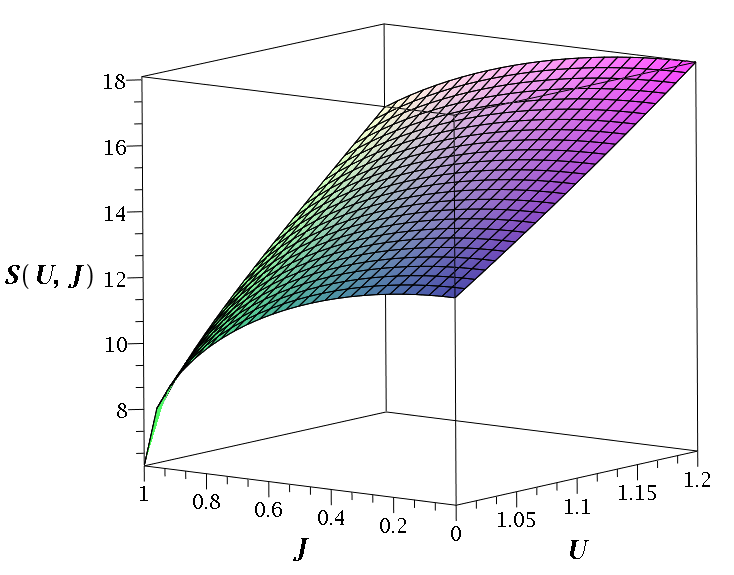} \\
\label{fig13}
\end{figure}
Figure 13.\\


\begin{thebibliography}{999}

\bibitem{Callen} H.B. Callen, Thermodynamics and an Introduction to Thermostatistics, 2nd ed. (John Willey \& Sons, Singapore, 1985) pp. 35, 183.
\bibitem{Greiner} W. Greiner, L. Neise and H. Stöker, Thermodynamics and Statistical Mechanics, 1st ed. (Springer-Verlag, NY, 1995) pp. 4, 218.
\bibitem{Dauxois} T. Dauxois, S. Ruffo, \textit{et. al.} (editors), Dynamics and thermodynamics of systems with long range interactions: an introduction, (Lect. Notes in Phys. vol 602), 1st ed. (Springer, Berlin, 2002). 
\bibitem{Johal} R.S. Johal, A. Planes and E. Vives, \textit{Preprint} cond-mat/0503329 (2005).
\bibitem{Campa} A. Campa, T. Dauxois and S. Ruffo, Phys. Rep. \textbf{480} 2009 57.
\bibitem{Bekenstein1} J.D. Bekenstein, Phys. Rev. D \textbf{7} (1973) 2333.
\bibitem{Bekenstein2} J.D. Bekenstein, Phys. Rev. D \textbf{12} (1975) 3077.
\bibitem{Hawking2} J.M. Baerdeen, B. Carter and S.W. Hawking, Commun. Math. Phys. \textbf{31} ( 1973) 161.
\bibitem{Candelas} P. Candelas and D.W. Sciama, Phys. Rev. Lett. \textbf{38} (1977) 1372.
\bibitem{Davies1} P.C.W. Davies, Rep. Prog. Phys. \textbf{41} (1978) 1313.
\bibitem{Wald} R.M. Wald, Liv. Rev. Rel. \textbf{4} (2001) 6.
\bibitem{Martinez1} E.A. Martinez, Phys. Rev. D \textbf{53} (1996) 7062.
\bibitem{Martinez2} E.A. Martinez, Phys. Rev. D \textbf{54} (1996) 6302.
\bibitem{Custodio} P.S. Custodio and J.E. Horvath, Am. J. Phys. \textbf{71 (12)} (2003) 1237.
\bibitem{York} J.W. York Jr, Phys. Rev. D \textbf{33} ((1986)) 2092.
\bibitem{Braden} H.W. Braden, J.D. Brown, B.F. Whiting and J.W. York Jr, Phys. Rev. D \textbf{42} (1990) 3376.
\bibitem{Brown} J.D. Brown, J. Creighton and R.B. Mann, Phys. Rev. D \textbf{50} (1994) 6394.
\bibitem{Lemos} C. Peça and J.P.S. Lemos, Phys. Rev. D \textbf{59} (1999) 124007.
\bibitem{Akbar} M.M. Akbar, Phys. Rev. D \textbf{82} (2010) 064001.
\bibitem{D'Inverno} R. D'Inverno, Introducing Einstein's Relativity, 1st ed. (Oxford University Press, NY, 1992) pp. 248.
\bibitem{Hawking4} S.W. Hawking, Phys. Rev. D \textbf{13} (1976) 191.
\bibitem{Bekenstein3} J.D. Bekenstein, Physics Today (Jan 1980) 24.
\bibitem{Misner} C.W. Misner, K. S. Thorne and J.A. Wheeler, Gravitation (W.H. Freeman, SF, 1973), pp. 878, 907.
\bibitem{Padmanabhan} T. Padmanabhan, Phys. Rep. \textbf{188} (1990) 285.
\bibitem{Oppenheim} J. Oppenheim, Phys. Rev. E \textbf{68} (2003) 016108.
\bibitem{Parker} B.R. Parker and R.J. McLeod, Am. J. Phys. \textbf{48 (12)} (1989) 1066.
\bibitem{Tishin} A.M. Tishin and Y.I. Spichkin, The Magnetocaloric Effect and its Applications, 1st ed. (Institute of Physics Publishing, Dirac House, UK, 2003) pp. 7
\bibitem{Davies2} P.C.W. Davies, Class. Quantum Grav. \textbf{6} (1989) 1909.
\bibitem{Koenig1} F.O. Koenig, J. Chem. Phys. \textbf{3} (1935) 29.
\bibitem{Stanley} H.E. Stanley, Introduction to Phase Transitions and Critical Phenomena, 1st ed. (Oxford University Press, London, 1971), pp. 34.

\end{thebibliography}
\end{document}